\documentclass[aps,pra,twocolumn,groupedaddress,showpacs,showkeys]{revtex4}

\usepackage{graphicx}
\begin{document}

\newcommand{\be}{\begin{equation}}
\newcommand{\ee}{\end{equation}}

\title{ Comment on " Low-frequency character of the Casimir force between metallic films".}

\author{ Giuseppe Bimonte}
\email[Bimonte@na.infn.it]
\affiliation{Dipartimento di Scienze Fisiche, Universit\`{a} di
Napoli Federico II, Complesso Universitario MSA, Via Cintia
I-80126 Napoli, Italy;\\ INFN, Sezione di Napoli, Napoli, ITALY\\
}

\date{\today}

\begin{abstract}
In Phys. Rev. {\bf E 70}, 047102 (2004), J.R. Torgerson and S.K.
Lamoreaux investigated for the first time the real-frequency
spectrum of   finite temperature correction to the Casimir force,
for metallic plates of finite conductivity. The very interesting
result of this study is that the correction from the TE mode is
dominated by low frequencies, for which the dielectric description
of the metal is invalid. However, their analysis of the problem,
based on more appropriate low-frequency metallic boundary
conditions, uses an incorrect form of boundary conditions for TE
modes. We repeat their analysis, using the correct boundary
conditions.   Our computations confirm their most important
result: contrary to the result of the dielectric model, the
thermal TE mode correction leads to an increase in the TE mode
force of attraction between the plates. The magnitude of the
correction has a value about twenty times larger than that quoted
by them.

\end{abstract}


\maketitle

In the recent literature on the Casimir effect, much attention has
been devoted to the issue of evaluating the corrections to the
Casimir force between metallic bodies, arising from the combined
effect of temperature and finite conductivity of the plates. An
estimate of these corrections \cite{sernelius}, using a dielectric
Drude model (with dissipation) for the plates, leads to
surprisingly large  deviations from the perfectly conducting case,
for separations among the plates greater than a micron or so, at
room temperature.  Several authors have criticized the validity of
these results, for different reasons. Torgerson and Lamoreaux
\cite{lamor}, in particular, performed for the first time a
spectral analysis of these thermal corrections along the {\it real
frequency axis},  while standard treatments based on Lifshitz
theory always deal with imaginary frequencies, which have a far
less clear physical meaning.   The new result of this very
interesting study is that the large corrections found in
\cite{sernelius} arise from {\it TE evanescent modes of  low
frequencies}. The frequencies involved are sufficiently low for
the dielectric description of the metal to be invalid. Torgerson
and Lamoreaux correctly suggest that a more realistic description
of the metal, in the frequency region of interest, can be obtained
in terms of Leontovich surface impedance boundary conditions
(b.c.). Following the  notations  of \cite{lamor}, we assume that
the plates surfaces are at $z=0$ and $z=a$. Then, for a TE mode of
frequency $\omega$, propagating along the $x$ axis,   the b.c. for
a good conductor read as: \be E_y=\pm \, \zeta \,
H_x\;,\label{ibc}\ee where the $+$ and $-$ sign refer to $z=0$ and
to $z=a$ respectively. For the surface impedance $\zeta$,
Torgerson and Lamoreaux use the following expression   \be
\zeta=(1-i) \sqrt{\frac{\omega}{8 \pi \sigma}} \ee    which is
valid for frequencies in the normal skin-effect region. However,
at this point Torgerson and Lamoreaux use an incorrect form of
fourth Maxwell's equation in vacuum, their Eq. (8), which does not
take account  of the $z$ component of the magnetic field. Indeed,
the magnetic field present in the empty gap between the plates can
be obtained from the second Maxwell equations \be
\vec{\nabla}\times \vec{E}-i\, \frac{\omega}{c}\, \vec{B}=0\;. \ee
For  TE modes, with ${\vec E}=\hat{y}\,E_y $, one obtains \be
\vec{B}= \frac{c}{\omega}\,\left(\hat{z}\,k\,E_y +
i\,\hat{x}\,\frac{\partial E_y}{\partial z}
\right)\;,\label{sm}\ee where $\vec{k}$ is the  transverse
wave-vector. We see that the magnetic field   has a $z$-component
$B_z=c \,k E_y /\omega$,  which was omitted in Eq. (8) of
\cite{lamor}, whose correct form really is (for $\mu$=1): \be
\frac{\partial H_x}{
\partial z}-\frac{\partial H_z}{\partial x}=  - \frac{i
\omega}{c}\,E_y\;.\ee

As a consequence of this error, the b.c. on the magnetic field
given in Eq. (9) of \cite{lamor} are incorrect either. In fact the
correct b.c. are best written in terms of the electric field. By
using the expression of $H_x$ in terms of $E_y$, obtained from Eq.
(\ref{sm}) above, we can rewrite the impedance b.c. for TE modes
  Eq. (\ref{ibc}) as \be E_y=\pm \,\frac{i \,
c\, \zeta}{\omega}  \, \frac{\partial E_y}{\partial z}\;.\ee If
one defines the spectrum $F_{\omega}$ of the thermal correction to
the Casimir force $F$ by the equation \be F= \frac{\hbar}{\pi^2 \,
c^3}\,\int_0^{\infty} d \omega \; F_{\omega}\;, \ee (attraction
corresponds to $F>0$) by simple computations analogous to those
after Eq. (9) of \cite{lamor}, one can get the following
expression for the TE-mode contribution to $F_{\omega}^{(TE)}$, in
the simple case of two identical plates: \be F_{\omega}^{(TE)}=
\omega^3 \, g(\omega)\,{\rm Re} \,\int_C p^2 dp
\left[\left(\frac{1+\zeta \,p}{1-\zeta \,p}\right)^2 e^{-2 i
\omega p a /c} -1\right]^{-1}\;, \label{te}\ee which should be
used in the place of Eq. (11) of \cite{lamor}. We note   that Eq.
(11) of \cite{lamor}  in fact reproduces, accidentally  the   TM
modes contribution to $F_{\omega}$: \be F_{\omega}^{(TM)}=
\omega^3 \, g(\omega)\,{\rm Re} \,\int_C p^2 dp
\left[\left(\frac{p+\zeta }{p-\zeta}\right)^2 e^{-2 i \omega p a
/c} -1\right]^{-1}\;. \label{tm}\ee Following Torgerson and
Lamoreaux, the integration path is separated into $C_1$ for $p=1$
to 0 (which describes the contribution from plane waves), and
$C_2$ with $p$ pure imaginary from $p=i \, 0$ to $i \, \infty$
(corresponding to evanescent waves). In our computations, we used
for $\sigma$ the expression \be \sigma(\omega)= \frac{ \sigma_0}{
1-i \,\tau \, \omega}\;,\ee with $\sigma_0= 3 \times 10^{17} \,
{\rm s}^{-1}$ and $\tau= 1.88 \times 10^{-14}$ s, which are the
values for Au. We  found that, both in the TM sector and in the
plane--wave TE sector, the spectra obtained from  Eqs. (\ref{te})
and (\ref{tm}) coincides, to a high degree of accuracy, with those
derived from dielectric b.c.. For $T=300$ K and $a=1 \mu$m, the
two approaches lead in these sectors to integrated forces  that
differ by a few parts in a thousand. Significant differences   are
found only in the evanescent TE sector.
 \begin{figure}
\includegraphics{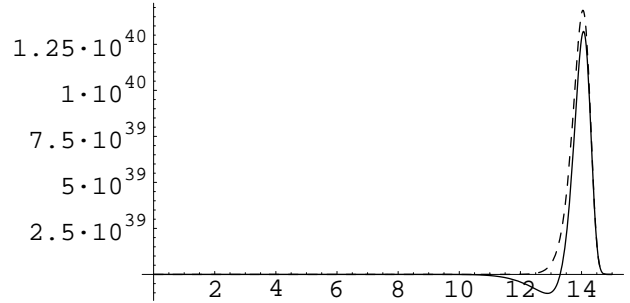}
\caption{\label{plotssigma}  Plots of  the contribution to
$F_{\omega}$ from the $C_1$ path (plane waves), for perfectly
conducting plates (dashed line) and for finite conductivity
boundary conditions, as functions of $\log_{10}(\omega)$. All are
for $a= 1 \; \mu$m, $T=300$ K. Treatment of the plates as
conducting metals fails above $\omega=10^{14}$ rad/s.}
\end{figure}
Results of numerical integration of Eq. (\ref{te}), for $a=1
\mu$m, $T=300$ K are shown in Figs. 1 and 2.  We see from Fig. 2
that the thermal correction from evanescent modes has a  {\it
positive} sign, which means that it represents an {\it attractive}
contribution, contrary to the result obtained from dielectric b.c.
(see Fig. 1 of \cite{lamor}), and  in agreement with what was
reported by Torgerson and Lamoreaux. The integrated force for the
$C_2$ path is  $38$ times larger than the $C_1$ integration, while
Torgerson and Lamoreaux reported a result only 1.47 times greater.
The total net force for both paths is  36.5 times  larger than the
perfectly conducting case, while the above authors obtained a
result 1.75 larger. As discussed in \cite{lamor} treatment of the
plates as good conductors is not valid above $\omega=10^{14}$
rad/s.
\begin{figure}
\includegraphics{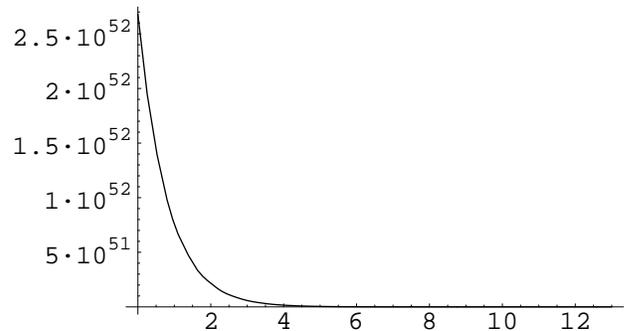}
\caption{\label{plotssigma} Plot of  the contribution to
$F_{\omega}$ from the $C_2$ path (evanescent waves), for  finite
conductivity boundary conditions, as function of
$\log_{10}(\omega)$,   for $a= 1 \; \mu$m, $T=300$ K.  The
integrated force is {\it attractive} and has a magnitude 38 times
larger than the $C_1$ integration. The total net force for both
paths is  36.5 times greater than the perfectly conducting case.
Treatment of the plates as conducting metals fails above
$\omega=10^{14}$ rad/s.}
\end{figure}

Our conclusion is that, despite the error in the b.c., the
qualitative  results of Ref.\cite{lamor} are correct: if one
models the plates as good conductors, one finds  that the TE mode
thermal correction leads to an increase in the TE mode force,
contrary to what is obtained from the dielectric model, and the
magnitude of the correction is  over thirtyfive   times larger
than the perfectly conducting case. Finally, we  remark that Ref.
\cite{lamor2} reports the same erroneous form of impedance b.c.
for the TE modes, as that of \cite{lamor}.

\end{document}